\documentstyle[12pt]{article}

\begin{document}
\title{
 Relativistic Acoustic Geometry
 }
\author{
Neven Bili\'c
 \\
 Rudjer Bo\v{s}kovi\'{c} Institute, 10001 Zagreb, Croatia \\
E-mail: bilic@thphys.irb.hr
}
\maketitle
\date{\today}
\begin{abstract}
 Sound wave propagation
 in a relativistic perfect fluid
with a non-homogeneous isentropic flow
 is studied in terms of
 acoustic geometry.
The sound wave equation turns out to be equivalent to
the equation of motion for a massless
scalar field propagating in a
curved space-time geometry.
The geometry is described by the
acoustic metric tensor that
depends locally on the
 equation of state and the four-velocity
 of the fluid.
 For a relativistic supersonic flow
 in curved space-time
 the ergosphere and acoustic horizon
 may be defined in a way analogous to
 the non-relativistic case.
 A general-relativistic expression for
 the acoustic analog of surface gravity
 has been found.
\end{abstract}
%

%
\section{Introduction}
\label{introduction}
Since
 Unruh's discovery \cite{unr81} that
 a supersonic flow may cause
  Hawking radiation, the
analogy between the theory of supersonic flow
and the black hole physics has been
extensively discussed \cite{jac91,unr95,bro,jac96,cor-jac96,vis}.
All the works have considered non-relativistic
flows in flat space-time.
In a comprehensive review \cite{vis} Visser defined the notions
of ergo-region, acoustic apparent horizon and acoustic
event horizon, and investigated the properties of various
acoustic geometries.

In this paper we study the acoustic
geometry for relativistic fluids moving in
curved space-time.
Under extreme conditions of very high density
and temperature,
the velocity of the fluid may be comparable
with the speed of light $c$.
In such conditions,
the speed of sound may also be close to $c$.
In particular, the equation of state of  an ideal
ultrarelativistic gas yields the speed of
sound as $c_s= c/\sqrt{3}$.
Such conditions are physically realistic in astrophysics,
early big-bang cosmology and relativistic
collisions of elementary particles and ions.

We restrict attention to perfect, isentropic fluids
and,
 in addition,
 we impose
that the flow be irrotational.
In most of our considerations we  make no assumption
about the space-time metric.
However, in the calculation  of surface gravity we
assume stationary space-time and steady flow.
Therefore, in our discussion we make no distinction between
acoustic
apparent and event horizons.

We organize the paper as follows.
In sections
\ref{acoustics}
and
\ref{metric}
we derive the relativistic acoustic wave equation
in terms of an acoustic metric tensor.
Section
\ref{horizon}
describes the notions of acoustic horizon and ergo-region.
In section
\ref{hawking}
we give the
calculation of surface gravity at the
horizon
and  also study the acoustic Unruh
effect and Hawking temperature.
We draw our conclusions in
 section
 \ref{conclusion}.
 In
appendix
\ref{basic}
we outline the basic
relativistic fluid dynamics
and derive a few  formulae
needed for our consideration.

\section{Relativistic Acoustics}
\label{acoustics}
We begin by deriving the sound wave equation
for a perfect, irrotational gravitating fluid.
We denote by
 $u_{\mu}$ ,
$p$, $\rho$, $n$   and $s$ the velocity, pressure,
energy density, particle number density
and entropy density of the fluid.
The basic equations of relativistic fluid dynamics
(for details see appendix \ref{basic})
needed for our discussion
are the continuity equation
\begin{equation}
\partial_{\mu}({\sqrt{-g}}\,  n u^{\mu})=0 \,
\label{eq202}
\end{equation}
and the equation of potential flow
\begin{equation}
  wu_{\mu}= -\partial_{\mu} \phi \, ,
\label{eq107}
\end{equation}
where the specific enthalpy $w$
is defined as
\begin{equation}
  w= \frac{p+\rho}{n} .
\label{eq207}
\end{equation}

Following the standard procedure
\cite{vis,lan}
   we linearize
equations (\ref{eq202}) and
(\ref{eq107})
by introducing
\begin{eqnarray}
w\rightarrow w+\delta w , & & n\rightarrow n+\delta n ,
\nonumber \\
u^{\mu}\rightarrow u^{\mu}+\delta u^{\mu},
 & & \phi\rightarrow \phi+\delta \phi,
\label{eq008}
\end{eqnarray}
where the quantities
$\delta w$,
$\delta n$,
$\delta u^{\mu}$
and $\delta \phi$
are small
acoustic disturbances
around
 some average bulk motion represented by
 $w$, $n$,
 $u^{\mu}$
and $\phi$.
For notational simplicity,
in our equations
we shall use
$\varphi$
instead of $\delta\phi$.
Fluctuations of the metric due to acoustic
disturbances will be neglected.
Note that the normalization
(\ref{eq101}) implies
\begin{equation}
g_{\mu\nu} u^{\mu}\delta u^{\nu}=0.
\label{eq109}
\end{equation}
Substituting (\ref{eq008})
into (\ref{eq107}) and retaining only
the terms of the same order of smallness,
we find
\begin{equation}
\delta w=-u^{\mu}\partial_{\mu}\varphi
\label{eq009}
\end{equation}
and
\begin{equation}
w \delta u^{\mu}=-g^{\mu\nu}\partial_{\nu}\varphi
+u^{\mu}
u^{\nu}\partial_{\nu}\varphi .
\label{eq010}
\end{equation}
Similarly, linearization of the continuity equation
(\ref{eq202}) gives
\begin{equation}
\partial_{\mu}({\sqrt{-g}}\,\delta n u^{\mu})+
\partial_{\mu}({\sqrt{-g}}\,  n \delta u^{\mu})=0.
\label{eq011}
\end{equation}
Replacing $\delta n$ by
\begin{equation}
\delta n= (\frac{\partial w}{\partial n})^{-1} \delta w ,
\label{eq012}
\end{equation}
and substituting (\ref{eq009}) and
(\ref{eq010}) into
(\ref{eq011}), we obtain the wave equation
\begin{equation}
\partial_{\mu}
\left\{
\frac{n}{w}
{\sqrt{-g}}\,
\left[ g^{\mu\nu}-\left( 1-
(\frac{n}{w} \frac{\partial w}{\partial n})^{-1} \right)
u^{\mu}
u^{\nu}\right]\right\} \partial_{\nu}\varphi =0.
\label{eq013}
\end{equation}
This equation describes the propagation of the linearized
field
$\varphi$.
The space-time dependence of the background
 quantities $n$, $w$ and $u_{\mu}=-w^{-1}\partial_{\mu}\phi$
 is constrained by
 the equations of motion
for a perfect, isentropic and irrotational gravitating fluid.

In  the simplest case of
   a homogeneous flow
  in flat space-time,
  equation (\ref{eq013}) becomes
  the free wave equation
\begin{equation}
(\partial_t^2 -
c_s^2
\Delta)\varphi=0,
\label{eq014}
\end{equation}
where the quantity $c_s$ is the sound wave velocity
given by
\begin{equation}
c_s^2=\frac{n}{w}\left. \frac{\partial w}{\partial n}
 \right|_{s/n}.
\label{eq015}
\end{equation}
The subscript $s/n$ denotes that
the derivative is taken while keeping the
specific entropy $s/n$ constant,
i.e.  for an isentropic process.
 By making use of equation
 (\ref{eq004})
 from appendix \ref{basic},
 it may be easily verified that this definition
 is equivalent to the standard definition of the
 speed of sound
  \cite{lan}
\begin{equation}
c_s^2=\left. \frac{\partial p}{\partial\rho}
 \right|_{s/n} .
\label{eq016}
\end{equation}

Introducing the symmetric tensor
\begin{equation}
f^{\mu\nu}=
\frac{n}{w}
{\sqrt{-g}}\,
\left[ g^{\mu\nu}-\left( 1-
c_s^{-2} \right)
u^{\mu}
u^{\nu}\right],
\label{eq116}
\end{equation}
we rewrite the wave equation (\ref{eq013}) as
\begin{equation}
\partial_{\mu}
f^{\mu\nu}
\partial_{\nu}\varphi=0.
\label{eq216}
\end{equation}
This form of the acoustic wave equation
 will be used to construct the acoustic metric.

At this stage  it is instructive to calculate
the tensor $f_{\mu\nu}$ in
the flat-metric non-relativistic limit.
To do this, we
reinstate the speed of light $c$ and
make the replacements
\begin{eqnarray}
w\rightarrow mc^2, & &
n\rightarrow \rho_{\rm NR}/m,
\nonumber \\
 g^{\mu\nu}
\rightarrow
 \eta^{\mu\nu},
 & &
u^{\mu}
\rightarrow (1,v^i/c),    \\
 c_s
\rightarrow v_s/c
 & &
\nonumber
\label{eq017}
\end{eqnarray}
in (\ref{eq116}). Retaining only the leading terms in
$1/c$, we find
\begin{equation}
f^{\mu\nu}
\rightarrow
f_{\rm NR}^{\mu\nu} =\frac{\rho_{\rm NR}}{m^2c^2v_s^2}
\left(\begin{array}{ccc}
c^2    &   &  c v^j   \\
       &   &          \\
c v^i &  &-\delta^{ij} v_s^2 +v^i v^j
\end{array} \right).
\label{eq018}
\end{equation}
This $4\times 4$ matrix,
 apart from the overall constant factor $1/(mc)^2$ and the
minus sign due to the
signature convention, precisely equals the
corresponding  non-relativistic
tensor
discussed by Unruh \cite{unr95} and Visser
\cite{vis}.

\section{Acoustic metric}
\label{metric}

We now show that
equation (\ref{eq216}) with
 (\ref{eq116})  can be put in the form
 of the d'Alembertian equation of motion for a
  massless scalar field
  propagating in a (3+1)-dimensional
  Lorentzian geometry described by the acoustic
  metric tensor $G_{\mu\nu}$.
  We proceed in a way analogous to
  \cite{vis}.

  We first introduce the
  acoustic metric tensor
$G_{\mu\nu}$ and its inverse
$G^{\mu\nu}$ such that
\begin{equation}
f^{\mu\nu} =\sqrt{-G}\, G^{\mu\nu} ,
\label{eq019}
\end{equation}
where
\begin{equation}
G={\rm det}(f^{\mu\nu}).
\label{eq020}
\end{equation}
These equations imply
\begin{equation}
\det(G_{\mu\nu})=
\det(G^{\mu\nu}))^{-1}=
 G.
\label{eq021}
\end{equation}
It follows from equations
(\ref{eq116}) and
(\ref{eq019}) that the metric
$G_{\mu\nu}$
must be of the form
\begin{equation}
G_{\mu\nu} =k
[g_{\mu\nu}-(1-c_s^2)u_{\mu}u_{\nu}] ,
\label{eq022}
\end{equation}
where $k$ is a conformal factor which will be
fixed later.
Indeed, one may easily check that
using this expression and
(\ref{eq019}) one obtains
\begin{equation}
G^{\mu\gamma}
G_{\gamma\nu} =k
\frac{n
\sqrt{-g}}{w
\sqrt{-G}}\,
\delta^{\mu}_{\nu} \, .
\label{eq023}
\end{equation}

To calculate the determinant of
$G_{\mu\nu}$,
it is convenient to use comoving
coordinates \cite{tau69}.
In the general comoving coordinate system,
the four-velocity vector is given by
\begin{equation}
u^{\mu}=
\frac{
\delta^{\mu}_0
}{\sqrt{g_{00}}}\,  ;
 \;\;\;\;\;\;\;
u_{\mu}=
\frac{
g_{\mu 0}
}{\sqrt{g_{00}}}\,
\label{eq024}
\end{equation}
and hence
\begin{equation}
G =k^4
\det \left(g_{\mu\nu}-(1-c_s^2)
\frac{
g_{\mu 0}
g_{\nu 0}
}{g_{00}}\right).
\label{eq025}
\end{equation}
This expression may be simplified
using standard algebraic properties of a
determinant. We find
\begin{equation}
G =k^4 c_s^2 g.
\label{eq026}
\end{equation}
This equation combined with  (\ref{eq023}) fixes
the factor $k$ as
\begin{equation}
k =\frac{n}{wc_s}\, .
\label{eq027}
\end{equation}

Thus, we have shown that the acoustic wave equation
may be written in the form
\begin{equation}
\frac{1}{\sqrt{-G}}\,
\partial_{\mu}
(\sqrt{-G}\,
G^{\mu\nu})
\partial_{\nu}\varphi=0,
\label{eq028}
\end{equation}
with the acoustic metric tensor given by (\ref{eq022})
and its inverse by
\begin{equation}
G^{\mu\nu} =\frac{1}{k}
\left[g^{\mu\nu}-(1-\frac{1}{c_s^2})u^{\mu}u^{\mu}
\right].
\label{eq029}
\end{equation}

\section{Ergosphere and acoustic horizon}
\label{horizon}

In analogy to the Kerr black hole
we may now define the {\em ergo-region} as
a region in space-time where
the stationary Killing vector
becomes space-like.
The magnitude $||\xi||$ of $\xi^{\mu}$
may be calculated in terms of the three
velocity $v$ defined in appendix
\ref{basic}.
Using equation
(\ref{eq022})
(and
 (\ref{eq130})-(\ref{eq330})
 from appendix \ref{basic})
 we find
\begin{equation}
||\xi||^2 \equiv G_{\mu\nu} \xi^\mu \xi^\nu =
k \xi^{\mu}\xi_{\mu}[ 1-(1-c_s^2)\gamma^2]=
k \xi^{\mu}\xi_{\mu}
\gamma^2 (c_s^2-v^2).
\label{eq034}
\end{equation}
This becomes negative when the magnitude of
the flow velocity $v$ exceeds the speed of sound $c_s$.
 Thus, as in non-relativistic acoustics \cite{vis},
 any region of supersonic flow is an ergo region.
 The boundary of the ergo-region is
 a hypersurface $\Sigma_{c_s}$
 defined by the equation
\begin{equation}
v^2-c_s^2=0.
\label{eq035}
\end{equation}
Since the stationary Killing field
 $\xi^{\mu}$ becomes null on
 $\Sigma_{c_s}$,
 this hypersurface is called {\em stationary limit surface}
 \cite{haw73}.
The hypersurface
 $\Sigma_{c_s}$
 may, in general, be quite complicated,
in particular if the fluid equation of state,
and hence the speed of sound,
is space-time dependent.
For simplicity,
in the following discussion  we assume
 $c_s=\rm const$  throughout
the fluid.

To define acoustic horizon
we use the concept of wave velocity
 of a hypersurface
\cite{tau78}.
Let $\{\Sigma_a\}$
denote a set of hypersurfaces defined by
\begin{equation}
v^2-a^2=0,
\label{eq135}
\end{equation}
where $a$ is a constant, $0\leq a <c$.
The stationary limit surface
$\Sigma_{c_s}$ is contained in
$\{\Sigma_a\}$.
Equation (\ref{eq135}) together with the equation
\begin{equation}
x_0={\rm const}
\label{eq235}
\end{equation}
determines a two-dimensional surface $\sigma_a$.
The normal
 to  each $\Sigma_a$ is given by the vector field
\begin{equation}
 n_{\mu}\propto
\partial_{\mu}v^2.
\label{eq335}
\end{equation}
 Now  we define
\begin{equation}
v_{\perp}= -\frac{n^{\mu}u_{\mu}}{
\sqrt{
(n^{\mu}u_{\mu})^2-n^{\mu}n_{\mu}}}
\label{eq036}
\end{equation}
as the {\em wave velocity} of two-dimensional surfaces
$\sigma_a$ as measured by the observer $u$,
whose world-line is such that
 the four-velocity $u_{\mu}$
 evaluated at a point of $\Sigma_a$
 is tangent to the world-line
 of $u$.
Clearly,
$v_\perp^2 \leq c^2$ if and only if $\Sigma_a$ is time-like or null,
i.e. if and only if
$n^{\mu}n_{\mu} \leq 0$.
If $\Sigma_a$ is time-like, $n_{\mu}$ may be normalized
  as $n^{\mu}n_{\mu}=-1$.
 In this case,
 we decompose
 $u^{\mu}$ as
\begin{equation}
u^{\mu}=
v_{\perp}
\gamma_{\perp} n^{\mu}+
\gamma_{\perp} L^{\mu} ,
\label{eq037}
\end{equation}
where
\begin{equation}
\gamma_{\perp} = \frac{1}{\sqrt{1-v_{\perp}^2}} ,
\label{eq137}
\end{equation}
  and
\begin{equation}
L^{\mu}=\frac{1}{
\gamma_{\perp}}
(g^{\mu\nu}+n^{\mu}n^{\nu})
u_{\nu}
\label{eq237}
\end{equation}
is a time-like unit vector which represents a displacement
in $\Sigma_a$ along the projection of $u^{\mu}$.
We can regard (\ref{eq037}) as a decomposition of the
fluid flow into normal and tangential components with respect
to $\Sigma_a$.
The tangential three-velocity
$v_{\|}^i$
may be found
 by decomposing the vector
 $L^{\mu}$
 in a way similar to
 (\ref{eq230})
 from appendix \ref{basic}:
\begin{equation}
L^{\mu}=
\gamma_{\|} t^{\mu}+
(g^{\mu\nu}-
t^{\mu}t^{\nu})
 L_{\nu} ,
\label{eq337}
\end{equation}
where
\begin{equation}
\gamma_{\|} = t_{\mu}L^{\mu}=
 \frac{1}{\sqrt{1-v_{\|}^2}} ,
\label{eq437}
\end{equation}
and
\begin{equation}
v_{\|}^i=\frac{1}{\gamma_{\|}} L^i\, ;
 \;\;\;\;\;\;\;
 i=1,2,3\, .
\label{eq537}
\end{equation}
For steady flow, $t^{\mu}n_{\mu}=0$, and hence equation
(\ref{eq037}), together with (\ref{eq330}) and (\ref{eq437}),
yields
\begin{equation}
\gamma=
\gamma_{\|}
\gamma_{\perp}.
\label{eq637}
\end{equation}

Now we define the {\em acoustic horizon} as a
hypersurface $\cal H$ defined by the equation
\begin{equation}
 \frac{(n^{\mu}u_{\mu})^2}{
(n^{\mu}u_{\mu})^2-n^{\mu}n_{\mu}}-c_s^2=0 ,
\label{eq038}
\end{equation}
i.e. a hypersurface the wave velocity of which
equals the speed of sound at every point.
The acoustic horizon $\cal H$ and the stationary limit surface
$\Sigma_{c_s}$  in general do not coincide.
Equation (\ref{eq637}) states that $\cal H$
is inside the ergo-region and
 overlaps with  $\Sigma_{c_s}$
 at those points where
$L^{\mu}=t^{\mu}$, i.e. at the points where the three
velocity $v^i$ is perpendicular to the-twosurface
$\sigma_{c_s}$.

Let us illustrate our general considerations by
an example. Consider a two-dimensional axisymmetric
flow in a stationary axisymmetric space-time.
Physically, this
 may be a model for
the equatorial slice of a rotating star,
or for a draining flow in
an axially symmetric bathtub.
In the coordinates
$t$, $\phi$, $r$,
the general form of
a stationary axisymmetric metric
is given by
\cite{wal}
\begin{equation}
ds^2=-V(r)(dt-W(r)d\phi)^2
-V(r)^{-1}r^2 d\phi^2
-\Omega(r)^2dr^2 ,
\label{eq055}
\end{equation}
where $V$, $W$  and $\Omega$
are arbitrary functions of $r$
approaching
$V,\Omega\rightarrow 1$
and
$W\rightarrow 0$ as
$r$ tends to infinity.
We further assume that these functions are
positive and non-singular in the region
occupied by the fluid.
According to equation
(\ref{eq031}),
the fluid velocity takes the form
\begin{equation}
u^{\mu}=
\gamma
 \left(
 V^{-1/2}
-W v_{\phi};v_{\phi},v_{r}
\right)\, ,
\label{eq057}
\end{equation}
where
\begin{equation}
v^2=
 V^{-1}r^2
v_{\phi}^2 +\Omega^2 v_{r}^2,
\label{eq058}
\end{equation}
with
 $v_{\phi}$ and $v_r$ being
the $r$-dependent
azimuthal and radial components,
respectively.
The surfaces  $\sigma_a$
 defined by equation (\ref{eq135}) with
(\ref{eq235}) are  concentric circles with the centre at
the origin.
The unit vectors $n^{\mu}$ and $L^{\mu}$
appearing in the decomposition of $u^{\mu}$
in (\ref{eq037})
 may now be easily constructed  using
(\ref{eq335}) and
(\ref{eq237}),
respectively.
We find
\begin{equation}
n^{\mu}=
\Omega^{-1}
 \left(0;0,1
\right)\, ,
\label{eq059}
\end{equation}
\begin{equation}
L^{\mu}=
\gamma_{\|}
 \left(
 V^{-1/2}
-W v_{\phi};v_{\phi},0
\right)\, ,
\label{eq060}
\end{equation}
where
\begin{equation}
v_{\|}^2=
 V^{-1}r^2
v_{\phi}^2 .
\label{eq061}
\end{equation}
From equation
(\ref{eq036}),
the wave velocity $v_{\perp}$ of
hypersurfaces $\sigma_a$ is
\begin{equation}
v_{\perp}= \gamma_{\|}
 \Omega v_r \, .
\label{eq062}
\end{equation}
The equation of continuity
and the condition that the flow be irrotational
constrain the radial dependence of
$v_{\perp}$ and $v_{\|}$.
From equation (\ref{eq202})
we find
\begin{equation}
n\gamma_{\perp}v_{\perp}r=c_1
\label{eq063}
\end{equation}
and similarly, as a consequence of
equation (\ref{eq006}), we obtain
\begin{equation}
w \gamma_{\perp}\gamma_{\|}\left(V v_{\|}-\frac{V^{1/2}W}{r}\right)
=c_2,
\label{eq064}
\end{equation}
 where  $c_1$ and
$c_2$ are arbitrary constants.
Of course, these expressions
are valid as long as
$v_{\perp}^2$ and
$v_{\|}^2$ do not exceed the speed of light squared $c^2=1$.
If we further assume that the particle number $n$
and the speed of sound $c_s$
are constant throughout
the fluid,
equation (\ref{eq063})
yields an explicit expression for
the horizon radius
\begin{equation}
r_{\rm H}=
\frac{
 \sqrt{1-c_s^2}}{c_s} \frac{|c_1|}{n} .
\label{eq065}
\end{equation}
The radius
of the stationary limit surface
can be found by solving
the equation
\begin{equation}
r=
\frac{
 \sqrt{1-c_s^2}}{c_s}
 \left[ \frac{c_1^2}{n^2} +
 \left(\frac{c_2}{w}+
\frac{V(r)^{1/2}W(r)}{
 \sqrt{1-c_s^2}}
\right)^2 V(r)
\right]^{1/2}.
\label{eq066}
\end{equation}
Evidently, the acoustic horizon coincides with
the stationary limit surface,
i.e. the boundary of the ergosphere,
 if both the azimuthal velocity
$v_{\phi}$ and the off-diagonal metric term $W(r)$ vanish.
It is remarkable that the horizon radius $r_{\rm H}$
does not depend on metric, whereas
the radius of
the ergosphere does.

\section{Acoustic surface gravity and Hawking temperature}
\label{hawking}

Having established the definition of the acoustic horizon,
we proceed to calculate surface gravity at the
horizon, restricting
attention to stationary
acoustic geometries.
If the flow is steady,
and the metric $g_{\mu\nu}$ is stationary,
 there exists a system of coordinates in which
 the components of $G_{\mu\nu}$
 are time independent and the
acoustic geometry is said to be {\em stationary}.

We shall define the surface
gravity  in terms of the Killing
field $\chi^{\mu}$ that is null on the horizon.
The surface gravity $\kappa$ may then be calculated
using
\cite{wal}
\begin{equation}
 G^{\mu\nu} \partial_{\nu}
 ||\chi||^2=
 -2\kappa \chi^{\mu}
\label{eq039}
\end{equation}
evaluated on the horizon,
where
\begin{equation}
 ||\chi||^2=
 G_{\alpha\beta} \chi^{\alpha}\chi^{\beta}.
\label{eq040}
\end{equation}
Since $||\chi||^2=0$ on the horizon, the derivative
 $\partial_{\mu}
 ||\chi||^2$ is normal to the horizon, so that
 on the horizon we have
\begin{equation}
 G^{\mu\nu}
 n_{\nu}\frac{\partial}{\partial n}
 ||\chi||^2 =
 2\kappa \chi^{\mu} ,
\label{eq041}
\end{equation}
where $\partial/\partial n\equiv n^{\mu}\partial_{\mu}$
 denotes the normal
derivative on the horizon.
The definition
of surface gravity by
equation (\ref{eq039}) is conformally invariant
\cite{jac93}.
Therefore, in the calculations that follow
 we drop the conformal factor
$k$ in
 $G_{\mu\nu}$.

First, consider the simplest case when
the stationary limit surface
$\Sigma_{c_s}$ and
 the acoustic horizon $\cal H$ coincide.
Then,
 $\chi^{\mu}$ is equal to
 the stationary Killing field $\xi^{\mu}$ which,
according to equation (\ref{eq034}),
is null on the horizon.
A straightforward calculation gives
\begin{equation}
 G^{\mu\nu} n_{\nu}   =
 n^{\mu}-\frac{1}{v\gamma}u^{\mu}=
 -\frac{1}{v}
  \frac{\xi^{\mu}}{\sqrt{\xi^{\nu}\xi_{\nu}}}
\label{eq042}
\end{equation}
on the horizon.
Equation
(\ref{eq041})
applied to $\xi^{\mu}$
together with
(\ref{eq042})
and (\ref{eq034}) yields
\begin{equation}
\kappa=\frac{
 \sqrt{\xi^{\nu}\xi_{\nu}}
 }{1-c_s^2}
\frac{\partial}{\partial n}
(v-c_s),
\label{eq043}
\end{equation}
where it is understood that
 the derivative and
 the quantity
 $\xi^{\nu}\xi_{\nu}$ are to be taken
 at the horizon.
The corresponding Hawking temperature
$T_{\rm H}=\kappa/(2\pi)$ represents
the temperature as measured by an observer near
 infinity.
 The locally measured temperature follows the
 Tolman law, i.e.
\begin{equation}
T=\frac{\kappa}{
 2\pi \sqrt{\xi^{\nu}\xi_{\nu}}
 },
\label{eq044}
\end{equation}
so that the local temperature approaches
\begin{equation}
T\rightarrow
 \frac{1}{1-c_s^2}
\frac{\partial}{\partial n}
(v-c_s)
\label{eq045}
\end{equation}
as the acoustic horizon
$\cal H$
is approached.
Thus, in this limit, equation
(\ref{eq044}) with (\ref{eq043})
corresponds to the flat space-time
Unruh effect.
This behaviour  is analogous  to the
Schwarzschild black hole
\cite{wal99} except that our
local temperature does not diverge as
one approaches  $\cal H$.

Next, we consider the situation when  $\cal H$ and
$\Sigma_{c_s}$ do not coincide.
 In this case,
the calculation of surface gravity  may,
in general,
be quite non-trivial.
As in non-relativistic acoustics \cite{vis},
there is no particular reason to expect
the acoustic horizon to be in general a Killing
horizon.
However, the matter may greatly simplify if we assume
a certain symmetry, such that an analogy
with the familiar stationary axisymmetric black hole
may be drawn.

We consider the acoustic geometry that
satisfies
       the following assumptions:
\begin{enumerate}
\item
The acoustic metric is stationary.
\item
The flow is symmetric under
the transformation that generates
a displacement on the horizon along the projection of
the flow velocity.
\item
The metric $g_{\mu\nu}$ is invariant to the
above transformation.
\end{enumerate}
The last two statements are equivalent to saying that
a displacement
 along the projection of the flow velocity on the horizon
is an isometry.
The axisymmetric geometry considered in section
\ref{horizon} is an example that fulfils the above
requirements.

Suppose  we have identified the acoustic horizon
as a hypersurface $\cal H$ defined by equation
(\ref{eq038}).
Consider the vector field
 $\L^{\mu}$ defined by
(\ref{eq237}).
 Its magnitude is given by
\begin{equation}
||L||^2 \equiv G_{\mu\nu} L^\mu L^\nu =
 [ 1-(1-c_s^2)\gamma_{\perp}^2]=
\gamma_{\perp}^2 (c_s^2-v_{\perp}^2)
\label{eq046}
\end{equation}
Hence, the vector $L^{\mu}$ is null on the horizon.
We  now construct a Killing vector $\chi^{\mu}$
in the form
\begin{equation}
\chi^{\mu}=\xi^{\mu} +\omega \psi^{\mu} ,
\label{eq047}
\end{equation}
where the Killing vector
$\psi^{\mu}$
is the generator of the isometry group
of displacements on the horizon,
and the constant
$\omega$
 is chosen such that
$\chi^{\mu}$ becomes
parallel to $L^{\mu}$
and hence null
on the horizon.
The quantity
$\omega$
is analogous to the horizon
angular velocity of the
axisymmetric black hole \cite{wal}.

We  use a special coordinate system in which
the Killing vector $\psi$ has the components
\begin{equation}
\psi^{\mu}=\delta^{\mu}_{\psi},
\label{eq048}
\end{equation}
where the subscript $\psi$ denotes the spatial coordinate
$x^{\psi}$ associated to the above-mentioned isometry.
The metric components $g_{\mu\nu}$ and the flow velocity field
are, by assumptions 2 and 3,
independent of $x^{\psi}$.
In this coordinate system, the vector $L^{\mu}$
takes the form
\begin{equation}
L^{\mu}
 =
\gamma_{\|} \left(
 \frac{1}{\sqrt{g_{00}}}
-\frac{g_{0\psi}v_{\psi}}{g_{00}};v_{\psi},0,0
\right) \, .
\label{eq049}
\end{equation}
Requiring
this vector to be parallel to $\chi^{\mu}$
on the horizon, we set
\begin{equation}
\chi^{\mu}=
\sqrt{\chi^{\nu}
\chi_{\nu}}
L^{\mu},
\label{eq050}
\end{equation}
with
\begin{equation}
\sqrt{\chi^{\nu}
\chi_{\nu}}=
\gamma_{\|}^{-1} \left(
 \frac{1}{\sqrt{g_{00}}}
-\frac{g_{0\psi}v_{\psi}}{g_{00}}
\right)^{-1}
\label{eq150}.
\end{equation}
This equation combined
 with (\ref{eq047}) determines
the ``angular velocity" of the horizon
\begin{equation}
\omega=
 \left(
 \frac{1}{\sqrt{g_{00}}}
-\frac{g_{0\psi}v_{\psi}}{g_{00}}
\right)^{-1}
v_{\psi}.
\label{eq051}
\end{equation}

The calculation of the surface gravity is now straightforward.
 The magnitude of $\chi^{\mu}$ in
the vicinity of the horizon
is given
from equation (\ref{eq046})
by
\begin{equation}
||\chi||^2 =
\gamma_{\perp}^2 (c_s^2-v_{\perp}^2)
  \chi^{\nu}\chi_{\nu}.
\label{eq052}
\end{equation}
Using the definition of $L_{\mu}$
(\ref{eq237})
and equation
(\ref{eq050}), we find
\begin{equation}
 G^{\mu\nu} n_{\nu}   =
 n^{\mu}-\frac{1}{v_{\perp}\gamma_{\perp}}u^{\mu}=
 -\frac{1}{v_{\perp}}
  \frac{\chi^{\mu}}{\sqrt{\chi^{\nu}\chi_{\nu}}}
\label{eq053}
\end{equation}
on the horizon.
Equation
(\ref{eq041}),
together with
(\ref{eq052})
and (\ref{eq053}), yields
\begin{equation}
\kappa=\frac{
 \sqrt{\chi^{\nu}\chi_{\nu}}
 }{1-c_s^2}
\frac{\partial}{\partial n}
(v_{\perp}-c_s),
\label{eq054}
\end{equation}
where the values of
 the derivative
 and the quantity
 $\chi^{\nu}\chi_{\nu}$ are to be taken
 at the horizon.
 The non-relativistic limit of this equation
 in flat space-time coincides with
 the expression for surface gravity derived by
 Visser \cite{vis}.

\section{Concluding Remarks}
\label{conclusion}
We have studied
the relativistic acoustics in curved space-time
in terms of an acoustic metric tensor.
The acoustic metric tensor involves
 a non-trivial explicit space-time dependence.
 Apart from that,
all the features of the non-relativistic acoustic
geometry
in flat space-time
 \cite{unr95,vis}
 have their relativistic counterparts.
We have shown that
the propagation of
sound in a general-relativistic perfect fluid
with an irrotational flow
may be described by a scalar d'Alembert
equation in a curved acoustic geometry.
The acoustic metric tensor is a matrix that
depends on the fluid equation of state and involves
 two tensors:
 the background space-time metric tensor $g_{\mu\nu}$
and the product of the fluid four-velocities $u_{\mu}u_{\nu}$.
We have then shown that if in a non-homogeneous flow
there exists a supersonic region  called
{\em ergo-region}, then an  acoustic event horizon forms.
We have discussed an example of axisymmetric geometry
where the acoustic horizon does not coincide with
the boundary of the ergo-region.
We have calculated the surface gravity
$\kappa$ using a conformally invariant
definition \cite{jac93} which involves
a properly defined Killing vector that is
null on the horizon.
An acoustic horizon emits Hawking radiation
of thermal phonons \cite{unr81} at the temperature
$T_{\rm H}=\kappa/(2\pi)$ as measured by an
observer near infinity.

We believe that the effects we have discussed may be of
physical interest in all those phenomena that involve
relativistic fluids under extreme conditions.
This may be the case in astrophysics, early cosmology
and ultrarelativistic heavy-ion collisions.

\appendix

\section{Basic Fluid Dynamics}
\label{basic}

Consider
 a perfect gravitating relativistic fluid.
We denote by
 $u_{\mu}$ ,
$p$, $\rho$, $n$  and $s$ the velocity, pressure,
energy density, particle number density and
entropy density of the fluid.
The energy-momentum tensor of a perfect fluid is
given by
\begin{equation}
T_{\mu\nu}=(p+\rho) u_{\mu}u_{\nu}-p g_{\mu\nu}   ,
\label{eq001}
\end{equation}
where
 $g_{\mu\nu}$ is the metric tensor
with the Lorentzian signature
$(+---)$. Hence, in this convention, we have
\begin{equation}
 u^{\mu}u_{\mu}=  g_{\mu\nu}
 u^{\mu}u^{\nu}=1.
\label{eq101}
\end{equation}

Our starting point is the continuity equation
\begin{equation}
(nu^{\mu})_{;\mu}=
\frac{1}{\sqrt{-g}}
\partial_{\mu}({\sqrt{-g}}\,  n u^{\mu})=0 ,
\label{eq002}
\end{equation}
and the energy-momentum
conservation
\begin{equation}
{T^{\mu\nu}}_{;\nu}=0.
\label{eq102}
\end{equation}
This equation  applied to  (\ref{eq001}) yields
 the relativistic Euler's equation \cite{lan}
\begin{equation}
(p+\rho)u^{\nu}u_{\mu;\nu}
+\partial_{\mu}p
+u_{\mu}u^{\nu}
\partial_{\nu}p  =0.
\label{eq003}
\end{equation}
Euler's equation may be further simplified if one
restricts consideration to
an isentropic flow.
A flow is said to be {\em isentropic} when the specific entropy
$s/n$ is constant, i.e. when
\begin{equation}
\partial_{\mu}(\frac{s}{n})=0 .
\label{eq103}
\end{equation}

A flow may in general have a non-vanishing
vorticity
$\omega_{\mu\nu}$
defined as
\begin{equation}
\omega_{\mu\nu}= h^{\rho}_{\mu}
h^{\sigma}_{\nu} u_{[\rho;\sigma]},
\label{eq203}
\end{equation}
where
\begin{equation}
h^{\mu}_{\nu} =
\delta^{\mu}_{\nu}-
u^{\mu}u_{\nu}
\label{eq303}
\end{equation}
is the projection operator
which projects an arbitrary vector in space-time
into its component in the subspace orthogonal to
$u^{\mu}$.
A flow with vanishing vorticity, i. e. when
\begin{equation}
\omega_{\mu\nu}= 0,
\label{eq403}
\end{equation}
 is said to be {\em irrotational}.
 In the following we assume that the flow is
 isentropic and irrotational.

As a consequence of
equation (\ref{eq103}) and the
thermodynamic identity
\begin{equation}
 dw=
 T  d(\frac{s}{n})+\frac{1}{n}dp,
\label{eq004}
\end{equation}
with
$w=(p+\rho)/n$ being the specific enthalpy,
equation (\ref{eq003})
simplifies to
\begin{equation}
 u^{\nu}(wu_{\mu})_{;\nu}
-\partial_{\mu}w=0.
\label{eq005}
\end{equation}
Furthermore, for an isentropic irrotational flow,
 equation (\ref{eq403}) implies \cite{tau78}
\begin{equation}
 (wu_{\mu})_{;\nu}
 -(wu_{\nu})_{;\mu}=0.
\label{eq006}
\end{equation}
In this case, we may introduce a scalar function
$\phi$ such that
\begin{equation}
  wu_{\mu}= -\partial_{\mu} \phi ,
\label{eq007}
\end{equation}
where the minus sign is chosen for convenience.
It may be easily seen that
the quantity $wu^{\mu}$
in the form (\ref{eq007})
satisfies equation (\ref{eq005}).
Solutions of this form are the relativistic analogue
of potential flow in non-relativistic fluid dynamics
\cite{lan}.

It is convenient to
  parameterize the components of the fluid
  four-velocity in terms of
  three-velocity components.
  To do this, we
   use the projection operator
$g_{\mu\nu}-
t_{\mu}t_{\nu}$,
which projects a vector into
the subspace orthogonal to
the time translation Killing vector
 $\xi^{\mu}=(1;\vec{0})$ ,
 where $t_{\mu}$ is the unit vector
\begin{equation}
t^{\mu}=
\frac{\xi^{\mu}}{\sqrt{\xi^{\nu}\xi_{\nu}}}=
\frac{
\delta^{\mu}_{0}
}{\sqrt{g_{00}}}\,  ;
 \;\;\;\;\;\;\;
t_{\mu}=
\frac{\xi_{\mu}}{\sqrt{\xi^{\nu}\xi_{\nu}}}=
\frac{
g_{\mu 0}
}{\sqrt{g_{00}}}\, .
\label{eq130}
\end{equation}
We split up the vector $u_{\mu}$ in  two parts: one parallel
 and the other orthogonal to
 $t_{\mu}$:
\begin{equation}
u_{\mu}=
\gamma t_{\mu}+
(g_{\mu\nu}-
t_{\mu}t_{\nu})
u^{\nu}  ,
\label{eq230}
\end{equation}
where
\begin{equation}
\gamma=
 t^{\mu}
 u_{\mu} \, .
\label{eq330}
\end{equation}
From (\ref{eq230})
with (\ref{eq130}) we find
\begin{equation}
u_0    = \gamma
 \sqrt{g_{00}} .
\label{eq430}
\end{equation}
Equations (\ref{eq230}), (\ref{eq330}) and the normalization
of $u_{\mu}$ imply
\begin{equation}
\gamma_{ij}u^i u^j= \gamma^2-1 ,
\label{eq530}
\end{equation}
where $\gamma_{ij}$ is the induced
three-dimensional spatial metric:
\begin{equation}
\gamma_{ij}=
\frac{g_{0i}g_{0j}}{g_{00}}-g_{ij}   \, ;
 \;\;\;\;\;\;\;
 i,j=1,2,3.
\label{eq033}
\end{equation}
Now, it is natural to introduce
the three-velocity $v^i$, so that
\begin{equation}
u^i = \gamma v^i,
\label{eq630}
\end{equation}
with its covariant components $v_i$ and the magnitude squareed $v^2$
given by
\begin{equation}
v_i=\gamma_{ij} v^j, \;\;\;\;\;\; v^2=v^i v_i .
\label{eq032}
\end{equation}
It follows from
 (\ref{eq530}),
 (\ref{eq630}) and
 (\ref{eq032}) that
\begin{equation}
\gamma^2= \frac{1}{1-v^2}.
\label{eq730}
\end{equation}
Since $u^{\mu}$ and $t^{\mu}$ are time-like unit vectors,
 a consequence of
 (\ref{eq230}) is that
 $\gamma \geq 1$ and hence $0 \leq v^2 < 1$.

 Equations (\ref{eq130}) -
 (\ref{eq032})
 may now be used to calculate
 other covariant and contravariant components
 of the fluid four-velocity.
 We find
\begin{eqnarray}
\nonumber \\
u^{\mu}
 \!&\!=\!&\!
\gamma \left(
 \frac{1}{\sqrt{g_{00}}}
-\frac{g_{0j}v^j}{g_{00}};v^i
\right)   ,      \nonumber \\
u_{\mu}
 \!&\!=\!&\!
\gamma \left(
 \sqrt{g_{00}};
\frac{g_{0i}}{\sqrt{g_{00}}}-v_i
\right) .
\label{eq031}
\end{eqnarray}

%
%
\subsection*{Acknowledgment}
 This work  was supported by
 the Ministry of Science and Technology of the
 Republic of Croatia under Contract
 No. 00980102.
\vspace{0.2in}

\end{document}